\documentclass[journal=achre4,manuscript=article]{achemso}

\usepackage[version=3]{mhchem} 
\usepackage[T1]{fontenc}       
\usepackage{color}

\newcommand{\RR}[0]{\boldsymbol{R}}
\newcommand{\kB}[0]{k_{\mathrm{B}}}
\newcommand{\ex}[1]{\mathrm{e}^{#1}}

\author{Jaime Agudo-Canalejo}
\affiliation{Rudolf Peierls Centre for Theoretical Physics, University of Oxford, Oxford OX1 3PU, United Kingdom}
\alsoaffiliation{Department of Chemistry, The Pennsylvania State University, University Park, Pennsylvania 16802, United States}
\author{Tunrayo Adeleke-Larodo}
\affiliation{Rudolf Peierls Centre for Theoretical Physics, University of Oxford, Oxford OX1 3PU, United Kingdom}
\author{Pierre Illien}
\affiliation{ESPCI Paris, UMR Gulliver 7083, 10 rue Vauquelin, 75005 Paris, France}
\author{Ramin Golestanian}
\email{ramin.golestanian@ds.mpg.de}
\affiliation{Max Planck Institute for Dynamics and Self-Organization (MPIDS), Am Fassberg 17, D-37077 G\"ottingen, Germany}
\alsoaffiliation{Rudolf Peierls Centre for Theoretical Physics, University of Oxford, Oxford OX1 3PU, United Kingdom}

\title{Enhanced diffusion and chemotaxis at the nanoscale}

\begin{document}

\renewcommand*{\thefootnote}{\fnsymbol{footnote}}

\renewcommand*\tocentryname{}
\begin{tocentry}

\includegraphics[width=1\linewidth]{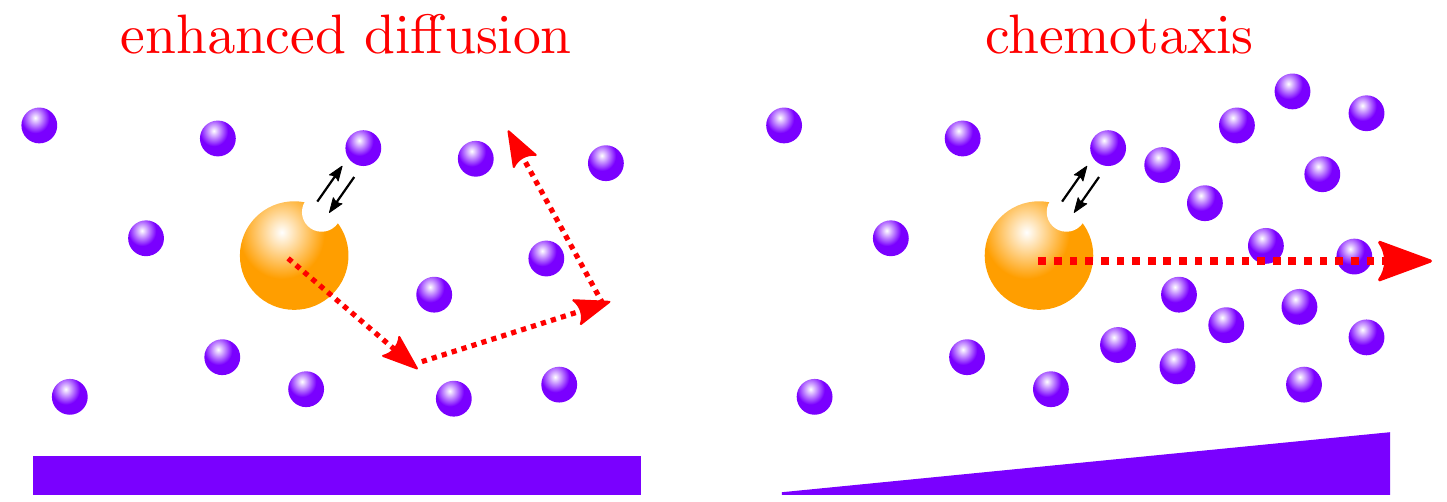}


\end{tocentry}

\begin{abstract}
Enzymes have been recently proposed to have mechanical activity associated with their chemical activity. In a number of recent studies, it has been reported that enzymes undergo enhanced diffusion in the presence of their corresponding substrate, when this substrate is uniformly distributed in solution. Moreover, if the concentration of the substrate is non-uniform, enzymes and other small molecules have been reported to show chemotaxis---biased stochastic movement in the direction of the substrate gradient---typically towards higher concentrations of this substrate, with a few exceptions. The underlying physical mechanisms responsible for enhanced diffusion and chemotaxis at the nanoscale, however, are still not well understood. Understanding these processes is important both for fundamental biological research, e.g. in the context of spatial organization of enzymes in metabolic pathways (metabolon formation), as well as for engineering applications, such as in the design of new vehicles for targeted drug delivery. In this Account, we will review the available experimental observations of both enhanced diffusion and chemotaxis, and we will discuss critically the different theories that have been proposed to explain the two. We first focus on enhanced diffusion, beginning with an overview of the experimental results. We then discuss the two main types of mechanisms that have been proposed, namely active mechanisms relying on the catalytic step of the enzymatic reaction, and equilibrium mechanisms which consider the reversible binding and unbinding of the substrate to the enzyme. We put particular emphasis on an equilibrium model recently introduced by us, which describes how the diffusion of dumbbell-like modular enzymes can be enhanced in the presence of substrate, thanks to a binding-induced reduction of the internal fluctuations of the enzyme. We then turn to chemotaxis, beginning with
an overview of the experimental evidence for the chemotaxis of enzymes and small molecules, followed by a description of a number of shortcomings and pitfalls in the thermodynamic and phenomenological models for chemotaxis introduced in those and other works in the literature. We then discuss a microscopic model for chemotaxis including both non-contact interactions and specific binding between enzyme and substrate recently developed by us, which overcomes many of these shortcomings, and is consistent with the experimental observations of chemotaxis. Finally, we show that the results of this model may be used to engineer chemically active macromolecules that are directed in space via patterning of the concentrations of their substrates.
\end{abstract}


\section{Introduction}

Nature has evolved to develop sophisticated mechanisms for microorganisms to propel themselves in low Reynolds number conditions and follow gradients of chemicals by coupling a sensing circuit to the motility machinery. In recent years, mimicking such capabilities in synthetic systems has been an important goal for nanotechnology. In particular, it is desirable to be able to make biocompatible nanoscale systems that can sense chemical gradients and move in response to them, so that they can be used for targeted drug delivery\cite{dey17}. Enzymes have been recently studied in this context, as they show great promise for such capabilities. Moreover, the motion of enzymes in response to chemical gradients may underly complex biochemical processes involving self-organisation at the molecular scale, such as metabolon formation. \cite{moll10,zhao17} A deeper understanding of the motion of enzymes may thus shed light on naturally-occurring metabolic pathways such as the Krebs cycle \cite{Wu2015}, as well as on ways to improve the design of synthetic metabolic pathways such as the CETCH cycle. \cite{schw16}

In this context, a number of recent studies have reported two types of behaviour that appear to be universal, in the sense that they occur for a wide range of enzymes with very different characteristics: \emph{enhanced diffusion}\cite{Muddana2010a,seng13,seng14,Riedel2015,Illien2017a,yu09} and \emph{chemotaxis}\cite{seng13,seng14,dey14,yu09,zhao17,guha17,jee17}. Enhanced diffusion refers to the observation that the diffusion coefficient of enzymes in a uniform solution of their substrate appears to increase with increasing concentration of this substrate. The diffusion coefficient can typically increase by a fraction of order unity, and this increase has been observed for all kinds of enzymes ranging from very fast, highly exothermic enzymes\cite{Riedel2015} to very slow, endothermic enzymes\cite{Illien2017a}. Chemotaxis, on the other hand, occurs when the concentration of the substrate is non-uniform, i.e.~in the presence of a substrate gradient. In this case, many different types of enzymes have been reported to move towards higher concentrations of their substrate.\cite{seng13,seng14,dey14,yu09,zhao17} In one study, however, enzymes have been reported to move towards lower concentrations of the substrate.\cite{jee17} Moreover, nanoscale chemotaxis has been observed not only for enzymes, but also for very small molecules such as molecular dyes.\cite{guha17}

There have been many recent developments towards a theoretical understanding of the physical mechanisms behind enhanced diffusion and chemotaxis. Here, we will review and critically discuss the theories that have been proposed for both enhanced diffusion and chemotaxis in light of the available experimental evidence, making particular emphasis on a set of related models recently introduced by us. \cite{Illien2017a,Illien2017b,agud18}

\section{Enhanced diffusion of enzymes}

\subsection{Experimental observations}

In a pioneering work, Muddana \latin{et al.} studied the diffusivity of enzyme molecules \latin{in vitro} and in dilute conditions using fluorescence correlation spectroscopy (FCS), and revealed that the diffusion coefficient of enzymes was enhanced when they were placed in the presence of substrate molecules\cite{Muddana2010a}. This first set of experiments was performed using urease, an enzyme known to catalyse an exothermic reaction (with a reaction enthalpy of $ \sim 20 \kB T$) with a fast turnover rate ($\sim 10^4$ reactions per second at substrate saturation). This phenomenon was reproduced with other enzyme molecules, such as catalase\cite{seng13}, or DNA polymerase\cite{seng14}. In all these measurements, the relative diffusion increase compared with the base value measured in the absence of substrate molecules could typically reach a few tens percent. Further studies have also shown how this intriguing phenomenon could play a role in biological self-organisation, for instance in the Krebs cycle metabolon formation\cite{Wu2015}.

In a first attempt to provide a physical mechanism to account for this phenomenon, Riedel \latin{et al.} compared the diffusivity of different enzyme molecules, and suggested that the relative diffusion enhancement could be correlated to the enthalpy of the reaction catalysed by the enzyme\cite{Riedel2015}.

Later on, a study involving the enzyme aldolase, which is known to catalyse an endothermic reaction and to have a very low turnover rate (at most 5 reactions per second at substrate saturation) revealed that, in spite of its singular kinetic and thermodynamic properties, it still displays enhanced diffusion\cite{Illien2017a}. More surprisingly, enhanced diffusion was also observed when this enzyme is placed in the presence of an inhibitor, that will only bind and unbind reversibly to the enzyme, suggesting that the catalytic step may not be necessary to observe enhanced diffusion.

\subsection{Models of enhanced diffusion \label{sec:endif}}

\begin{figure}
\centering
\includegraphics[width=0.8\linewidth]{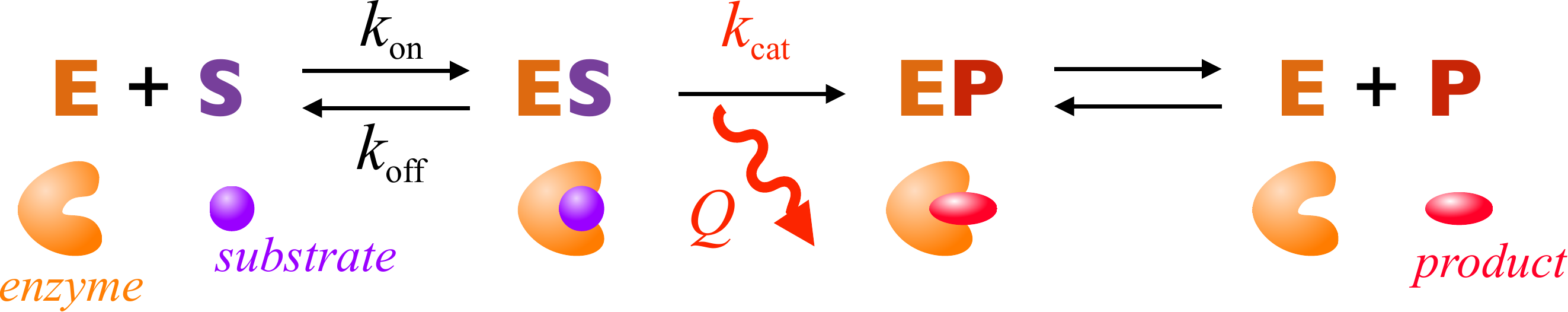}
\caption{Simplified Michaelis-Menten description of the chemical cycle: a substrate molecule reversibly binds to the active site of the enzyme with rates $k_\text{on}$ and $k_\text{off}$, and is converted into a product molecule with rate $k_\text{cat}$. The product is eventually released in the environment.}
   \label{MMcycle}
\end{figure}

In the light of these experimental results, the physical mechanism responsible for enhanced diffusion was to be elucidated. When placed in the presence of their substrate, enzymes explore a complex mechanochemical cycle, during which substrate molecules reversibly bind to the active site of the enzyme, and are subsequently transformed into product molecules that are eventually released in the solution. These steps are associated with conformational changes of the enzyme. Although a typical enzyme cycle may involve numerous steps and reaction intermediates, it can be reduced to a minimal description. We present in Figure~\ref{MMcycle} the simplest Michaelis-Menten description of enzyme kinetics: binding and unbinding of the substrate to the enzyme occur with rates $k_\text{on}$ and $k_\text{off}$, and the catalytic step, which occurs with rate $k_\text{cat}$, is typically associated with a heat transfer $Q$.

Relying on the initial experimental observations, that were associated to fast and exothermic enzymes, different physical mechanisms relying on the nonequilibrium or catalytic part of the chemical cycle were put forward. First, Riedel \latin{et al.} proposed a mechanism that relies on the idea that the heat released at each catalytic turnover is converted into an anisotropic compression of the enzyme and a translational boost\cite{Riedel2015}. This model was later criticised, as it relies on an underestimate of the friction coefficient of the enzyme and assumes that the released energy is partitioned over a small number of degrees of freedom\cite{Golestanian2015}. The effect of collective heating of the reaction sample due to heat release at each catalytic turnover was also investigated, and found capable of contributing significantly to enhanced diffusion for enzymes that are sufficiently fast or exothermic\cite{Golestanian2015}, although no evidence of collective heating was found in experiments with urease. \cite{dey15}  Enzymes were also described as nanoscale swimmers, and the effect of stochastic conformational changes triggered by  catalytic events was investigated, and shown to contribute to enhanced diffusion with a typical change in diffusivity of the order of $\Delta D \sim R^2 k_\text{cat} {c}/{(K + c)} $, where $R$ is the amplitude of the conformational changes\cite{Bai2015,Golestanian2015}. More recently, relying on experimental evidence for the existence of ballistic steps in enzyme trajectories, Jee \latin{et al.} have put forward a run-and-tumble description of the enzyme dynamics, and proposed an estimate of the resulting diffusion enhancement in relation with the catalytic rate $k_\text{cat}$\cite{jee17}. For all of these mechanisms, the typical diffusion coefficient of the enzyme in the presence of substrate molecules can be shown to take the generic form
\begin{equation}
\label{ }
D(c) = D_0 + \ell^2 k_\mathrm{cat} \frac{c}{K + c}
\end{equation}
where $\ell$ is a characteristic lengthscale of the active process, and $K$ the Michaelis constant.

We must also note that collective effects were considered by Mikhailov and Kapral, who described the enzyme solution as a collection of force dipoles, which have random amplitudes with non-Gaussian fluctuations coming from the nonequilibrium nature of catalysis, and which are coupled through hydrodynamic interactions\cite{Mikhailov2015}. The resulting fluctuations of the velocity field in the solvent can then yield enhanced diffusion of any object present in the solution. However, such an effect is controlled by the volume fraction of enzymes in the solution, which is typically very small in the FCS experiments. 

The experiments performed on the slow and endothermic enzyme aldolase constituted a strong indication that the catalytic step of the cycle may not be necessary to observe enhanced diffusion and led to the introduction of a new paradigm\cite{Illien2017a,Illien2017b}. Considering that binding and unbinding of substrate molecules may occur at rates much higher than catalysis ($k_\text{on}, k_\text{off} \gg k_\text{cat}$), we recently proposed a two-state model, where the enzyme is either free or bound, and where the catalytic events where substrate molecules are converted into product molecules occur at rates sufficiently small to be neglected.  Within this new picture, the typical diffusion coefficient of the enzyme takes the form
\begin{equation}
D(c) = D_0 + \Delta D \; \frac{c}{K + c}
\label{eq:Deq}
\end{equation}
where $\Delta D$ is the difference of diffusion coefficients in the free and the bound state. In order to understand the change in diffusivity induced by substrate binding and by the resulting changes in conformational fluctuations, we also proposed a simplified description of the internal degrees of freedom of the enzyme using a generalised dumbbell model that accounts for hydrodynamic interactions and that reveals how the internal fluctuations of the enzyme affect its overall diffusivity. We describe this model in detail in the following section.

In summary, the physical mechanisms that have been deduced from and related to the experimental observations can be sorted into two main categories: (i) enhanced diffusion can be related to the nonequilibrium step of the chemical cycle, and is controlled by the catalytic rate and/or the amount of heat released during a turnover; (ii) enhanced diffusion can be explained within an equilibrium picture, and originates from the fact that enzymes could diffuse significantly faster when they are bound rather than free. These two classes of models are in no way incompatible, and could jointly contribute to diffusion enhancement with different relative importance, depending on the kinetic, thermodynamic and structural properties of the enzymes studied experimentally.

\subsection{Equilibrium model of enhanced diffusion \label{sec:eqendif}}

Enzymes perform their function under conditions dominated by thermal fluctuations and viscous hydrodynamics, and are known to undergo conformational fluctuations during the catalytic cycle which affect their transport properties. The diffusion coefficient measured in fluorescence correlation spectroscopy experiments\cite{Muddana2010a,seng13,seng14,Riedel2015,Illien2017a} is an average over these conformations. In this section, we describe the \emph{dumbbell model} for enhanced diffusion, with which we have studied\cite{Illien2017a,Illien2017b} the effect of conformational fluctuations and of hydrodynamic interactions on the diffusion properties of modular macromolecules such as enzymes.

\begin{figure}
\centering
\includegraphics[width=0.6\linewidth]{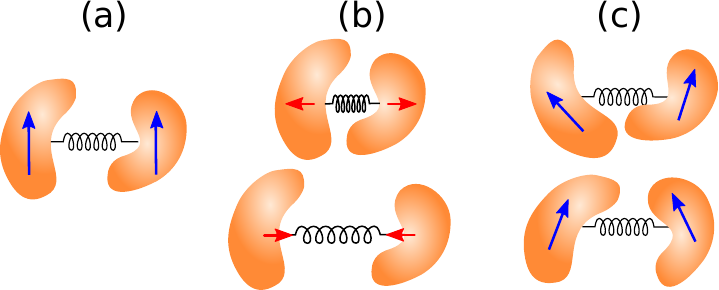}
\caption{A generalized dumbbell, made of two geometrically different subparts joined by an interaction potential (as represented by a spring), is used as a minimal model to capture the modularity and internal degrees of freedom of a real enzyme. The dumbbell can fluctuate around its equilibrium conformation (a), undergoing compressional (b) and orientational (c) fluctuations.}
   \label{dumbbel}
\end{figure}

A generic macromolecule is represented by an asymmetric dumbbell, so as to capture with a minimal number of ingredients the modularity, complex shape, and fluctuating internal degrees of freedom of real macromolecules; see Figure~\ref{dumbbel}. The generalized dumbbell consists of a pair of asymmetric Brownian particles, whose positions and orientations undergo thermal fluctuations, and are coupled through hydrodynamic interactions and an interaction potential.
Within this framework, we find that thermal fluctuations lead to an interplay between the internal and external degrees of freedom, resulting in negative fluctuation-induced corrections to the diffusion coefficient of the dumbbell. The effective diffusion coefficient has the generic form
\begin{equation}
D_{\text{eff}}=D_{\text{ave}}- \delta D_{\text{fluc}}
\label{eq:Dfluc}
\end{equation}
where the first term is the thermal average of contributions from translational modes, and the second term with $\delta D_{\text{fluc}}>0$ is due to internal compressional and rotational fluctuations. The fluctuation-induced correction is controlled by the asymmetry of the dumbbell, vanishing precisely for a symmetric dumbbell.\cite{Illien2017b} There is a crossover time between $D_{\text{ave}}$ and $D_{\text{eff}}$ which is the time it takes for the compressional mode to relax to equilibrium.\cite{Illien2017b}

In an
enzyme, the presence of substrate molecules activates the catalytic cycle, during which an enzyme is either free, or bound to a substrate or product molecule, and hence undergoes conformational fluctuations about different equilibrium states, see Figure~\ref{MMcycle}. Assuming that the nonequilibrium catalytic step is substantially slower than the conformation changes\cite{Rago2015}, there is a separation of time-scales which allows us to neglect the nonequilibrium step of the reaction, leaving an equilibrium description, involving just the binding and unbinding events.\cite{Illien2017a,Illien2017b} The effective diffusion coefficient of an enzyme will be affected by binding and unbinding through changes in the fluctuations of the internal degrees of freedom.

In the bound state, the enzyme is expected to have reduced fluctuations, which results in a decrease of the fluctuation-induced contribution $\delta D_{\text{fluc}}$ to the effective diffusion coefficient (\ref{eq:Dfluc}), and consequently in an enhancement of this effective diffusion coefficient. There are many possible mechanisms that could result in a reduction of fluctuations in the bound state. For example, in aldolase, binding brings the molecular structure closer together by a few angstroms\cite{Rago2015}, which is equivalent to reducing the equilibrium separation of the subunits in the dumbbell. It is also likely that the compressional fluctuations of the enzyme are reduced, as the interaction potential between the subunits becomes stiffer on binding, due to the presence of a substrate or product molecule. Binding will also affect the fluctuations in the orientations of the subunits, for example, through the closing of a hinged ``flap'' that is associated with the binding site\cite{Roberts2012}. These possible contributions to the change in the diffusion coefficient are estimated in Ref\citenum{Illien2017b}.

\begin{figure}
\centering
\includegraphics[width=0.4\linewidth]{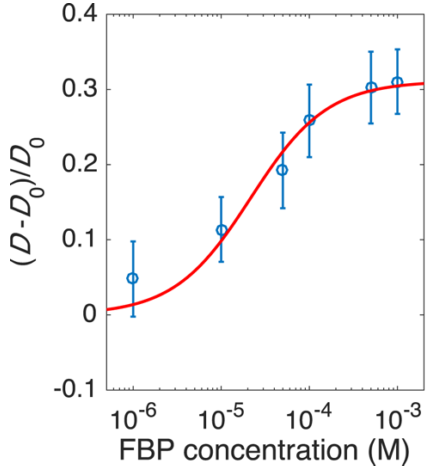}
\caption{Enhanced diffusion of the slow, endothermic enzyme aldolase in the presence of its substrate FBP. The experimental results (open circles) are well fitted (red line) to the equilibrium model given by (\ref{eq:Deq}), with the Michaelis constant $K$ extracted from the fit being comparable to that obtained from independent measurements. Reproduced with permission from Ref~\citenum{Illien2017a}. Copyright (2017) ACS Publications.  }
   \label{fig:aldolase}
\end{figure}

After taking an appropriate average of the free and bound conformations, we finally find that the diffusion coefficient of the enzyme shows a Michaelis-Menten-like dependence on the substrate concentration as in (\ref{eq:Deq}). The change in diffusion coefficient is positive $\Delta D > 0$, representing the increased diffusion coefficient of the bound state with respect to the free state, due to the reduction of the negative fluctuation-induced corrections to the diffusion coefficient in the bound state. In Ref\citenum{Illien2017a}, we applied this theory to the enhanced diffusion of aldolase in the presence of its substrate. A fit of (\ref{eq:Deq}) to the experimental measurements is shown in Figure~\ref{fig:aldolase}.

\section{Chemotaxis of enzymes and small molecules}

\subsection{Experimental observations}

\begin{figure}
\centering
\includegraphics[width=0.6\linewidth]{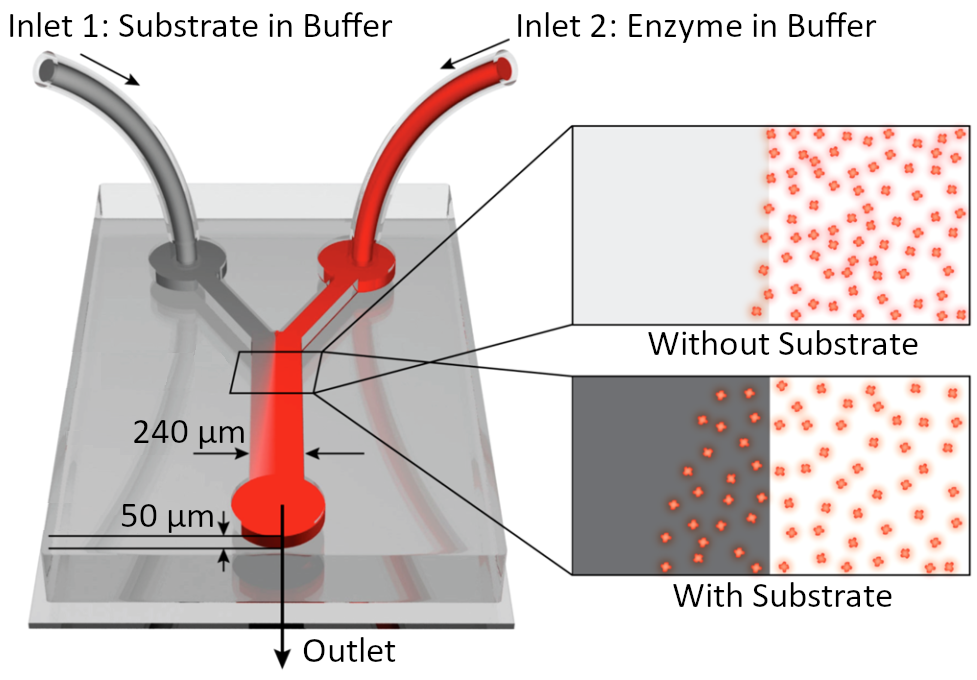}
\caption{Typical experimental observation of chemotaxis. (Left) The enzyme is introduced in one channel, whereas the substrate may or may not be introduced in the other channel. (Right) The fluorescence intensity of the enzymes is observed a certain distance down the channel. Compared with the control case with only buffer, the fluorescence profile shifts further to the left in the presence of substrate, indicating chemotactic behavior. Adapted with permission from Ref~\citenum{seng13}. Copyright (2013) ACS Publications.} 
   \label{fig:2inlet}
\end{figure}

The chemotactic behavior of catalytic enzymes has only been reported quite recently.\cite{yu09,seng13,dey14,seng14,dey15,jee17,zhao17,jose17} Experiments are typically performed using microfluidic devices with two or more inlets such as the one shown in Figure~\ref{fig:2inlet}. In such an experiment, a gradient of substrate is generated by introducing the substrate in only one of the inlets. By comparing the concentration profile of the enzyme a certain distance downstream in the presence of the substrate against the control experiment in the absence of it, one can ellucidate whether a chemotactic shift of the enzyme concentration towards (or away from) the substrate takes place. Typical shifts observed in such microfludic experiments imply chemotactic velocities of the order of several micrometers per second. \cite{seng13} 

Most experimental observations so far have shown chemotaxis \emph{towards} higher concentrations of the substrate: this includes RNA polymerase,\cite{yu09} catalase and urease,\cite{seng13,dey15} DNA polymerase,\cite{seng14} polymersomes encapsulating glucose oxidase,\cite{jose17} as well as hexokinase, phosphoglucose isomerase, phosphofructokinase and aldolase.\cite{zhao17} However, a recent publication reports chemotaxis \emph{away from} higher concentrations of the substrate for urease and acetylcholinesterase.\cite{jee17} These new results, at least in the particular case of urease, are in apparent conflict with the older results in the literature. We will show below that this contradiction may be explained by the presence of two competing mechanisms that dominate at different substrate concentrations. Chemotactic behavior has also been observed for molecular dyes towards a polymer to which they bind reversibly, \cite{guha17} suggesting that binding-unbinding alone (without catalysis) may be sufficient to induce chemotaxis.

\subsection{Thermodynamic and phenomenological models \label{sec:pheno}}

A thermodynamic theory describing chemotaxis as a consequence of specific binding to a solute was recently introduced by Schurr \latin{et al.}~\cite{schu13}, and has been used as support in several experimental studies\cite{guha17,zhao17} However, some points of concern regarding the use of this theory should be noted: (i) it is based on purely thermodynamic considerations, without consideration of thermal fluctuations or the hydrodynamics in the interfacial region (an approach known to be deficient\cite{ande89}); (ii) it is derived for colloids with a large number of binding sites $N \gg 1$, and therefore should not apply to enzymes or molecules whith $N\simeq1$; (iii) the predicted strength of chemotaxis is much weaker than experimentally observed;\cite{zhao17} (iv) it provides no link between enhanced diffusion and chemotaxis, while experimental evidence suggests they may be related;\cite{hong07,yu09,seng13,dey14,dey15,seng14,butl15,jee17,zhao17} and (v) it predicts that chemotaxis is always directed towards the substrate, in disagreement with recent experimental observations.\cite{jee17}

Other works\cite{seng13,jee17,hong07,dey14,weis17} in the literature have tried to understand the chemotaxis of enzymes (or other `active' particles) by phenomenologically linking the enhancement in diffusion as a function of substrate (or `fuel') concentration observed experimentally to the chemotactic behavior in response to gradients of this substrate. More precisely, suppose that the dependence of the diffusion coefficient of an enzyme on the concentration of its substrate $D(c_\mathrm{s})$ is known, from experiments at uniform substrate concentration. If the substrate concentration is now made position-dependent, $c_\mathrm{s}(\RR)$, it is then tempting to try to describe the motion of the enzyme by considering a position-dependent diffusion coefficient $D(\RR) \equiv D(c_\mathrm{s}(\RR))$. One should, however, be extremely careful when dealing with such a model, because the diffusion equation (or more generally, the Fokker-Planck equation) for a system with position-dependent diffusion coefficient is not uniquely defined, due to the presence of \emph{multiplicative noise}.\cite{lau07}

The problem of multiplicative noise is related to the following simple issue: given a position-dependent diffusion coefficient $D(\RR)$, how should we write the diffusion equation? Two possibilities that may come to mind are $\partial_t c_\mathrm{e} = \nabla^2 [D(\RR) c_\mathrm{e}]$ and $\partial_t c_\mathrm{e} = \nabla [D(\RR) \nabla c_\mathrm{e}]$. Notice that, while both forms are identical in the case of a constant $D$, they are different from each other if the diffusion coefficient is position-dependent! In fact, careful analysis\cite{lau07} shows that a whole family of diffusion equations can be derived from the same underlying Langevin dynamics, namely the family $\partial_t c_\mathrm{e} = \nabla [D(\RR) (\nabla c_\mathrm{e}) + (1-\alpha) (\nabla D(\RR)) c_\mathrm{e}]$, with $0 \leq \alpha \leq 1$. A given value of $\alpha$ corresponds to a different choice for the treatment of the multiplicative noise. In particular, the two simple forms of the diffusion equation quoted before correspond to $\alpha=0$ (It\^o choice) and $\alpha=1$ (anti-It\^o choice), respectively. The choice $\alpha=1/2$ corresponds to the Stratonovich treatment of the noise. The main problem faced by such phenomenological models is then: which diffusion equation is the correct one for a particular physical system? Unfortunately, there is no simple answer: the ambiguity can only be resolved by including more physical details into the model\cite{vanKampen1981,Schnitzer1993}. Below, we present a microscopic model of chemotaxis for which enhanced diffusion and a correct form of the diffusion equation arise naturally.

\subsection{A microscopic model of molecular chemotaxis}

Recently, we introduced a microscopic, first-principles model that is consistent with all the experimental observations of chemotaxis at the nanoscale described above. \cite{agud18} Starting from the fundamental properties of the enzyme (or small molecule), we take into account non-contact (e.g. van der Waals, electrostatic...) and hydrodynamic interactions between the enzyme and the substrate, as well as the possibility of complex formation through specific binding at a well-defined binding pocket of the enzyme. The non-contact interactions lead to a diffusiophoretic mechanism for chemotaxis, whereas binding leads to a novel type of chemotactic mechanism intimately related to enhanced diffusion.

\begin{figure}
\centering
\includegraphics[width=1\linewidth]{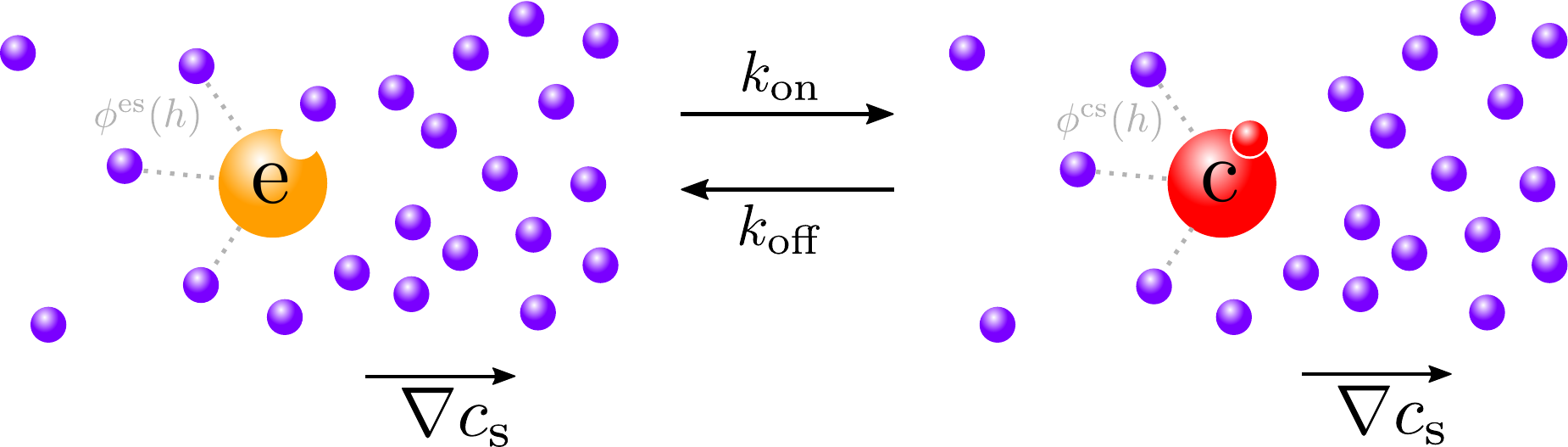}
\caption{Microscopic model for chemotaxis. The free enzyme (yellow) is in a gradient of substrate molecules (purple), with concentration $c_\mathrm{s}(\RR)$. The enzyme can bind one-on-one to a substrate molecule to form a complex (red), with binding rate $k_\mathrm{on}$ and unbinding rate $k_\mathrm{off}$, and also interacts with all other substrate molecules around it through a non-contact potential given by $\phi^\mathrm{es}(h)$ in the free state and $\phi^\mathrm{cs}(h)$ in the bound state.}
   \label{fig:chemotaxis}
\end{figure}

Consider an enzyme in a bath of substrate molecules with a concentration profile $c_\mathrm{s}(\RR)$, see Figure~\ref{fig:chemotaxis}. The enzyme can bind one-to-one to a substrate molecule to form an enzyme--substrate complex, with binding rate $k_\mathrm{on}$ and unbinding rate $k_\mathrm{off}$, but also interacts with all other substrate molecules through a non-contact potential $\phi(h)$   given by $\phi^\mathrm{es}(h)$ in the free state and $\phi^\mathrm{cs}(h)$ in the bound state, where $h$ is the distance between the surface of the enzyme and the substrate. The concentrations of free enzyme and of enzyme--substrate complex are denoted by $c_\mathrm{e}$ and $c_\mathrm{c}$, respectively. Starting from a microscopic description involving the Smoluchowski equation for all particles in the system, we showed \cite{agud18} that the total concentration of enzyme $c_\mathrm{e}^\mathrm{tot} = c_\mathrm{e} + c_\mathrm{c}$, both free and bound, is governed by the evolution equation
\begin{eqnarray}
\partial_t c_\mathrm{e}^\mathrm{tot} (\RR;t) = \nabla\cdot \left\{ D(\RR)\cdot\nabla c_\mathrm{e}^\mathrm{tot} - [\boldsymbol{V}_\mathrm{ph}(\RR)+\boldsymbol{V}_\mathrm{bi}(\RR)] c_\mathrm{e}^\mathrm{tot} \right\}.
\label{eq:totalevol}
\end{eqnarray}

The first term on the right hand side of (\ref{eq:totalevol}) is a diffusive term with a position-dependent diffusion coefficient given by
\begin{eqnarray}
D(\RR) = D_\mathrm{e} + (D_\mathrm{c} - D_\mathrm{e}) \frac{c_\mathrm{s}(\RR)}{K + c_\mathrm{s}(\RR)}
\label{eq:Deff}
\end{eqnarray}
where $D_\mathrm{e}$ and $D_\mathrm{c}$ are the (constant) diffusion coefficients of the free enzyme and the complex, and $K \equiv k_\mathrm{off}/k_\mathrm{on}$. This diffusion coefficient displays the same Michaelis--Menten-like dependence on the substrate concentration as the equilibrium model for enhanced diffusion described above.

The second term on the right hand side of (\ref{eq:totalevol}) represents the chemotactic drift, and involves two distinct contributions to chemotaxis. The first contribution corresponds to the phoretic velocity $\boldsymbol{V}_\mathrm{ph}(\RR)$ due to non-contact interactions, given by
\begin{eqnarray}
\boldsymbol{V}_\mathrm{ph}(\RR) = \boldsymbol{v}_\mathrm{ph,e}(\RR)+ [\boldsymbol{v}_\mathrm{ph,c}(\RR) - \boldsymbol{v}_\mathrm{ph,e}(\RR)] \frac{c_\mathrm{s}(\RR)}{K + c_\mathrm{s}(\RR)}
\label{eq:Vpheff}
\end{eqnarray}
where $\boldsymbol{v}_\mathrm{ph,e}(\RR)$ and $\boldsymbol{v}_\mathrm{ph,c}(\RR)$  correspond to the usual phoretic velocities \cite{derj47,ande89} of the free enzyme and the complex with
\begin{equation}
\boldsymbol{v}_{\mathrm{ph},i}(\RR) = \frac{\kB T}{\eta} \lambda_i^2 \nabla c_\mathrm{s}(\RR)~~\text{with}~~\lambda_i^2 \equiv \int_0^\infty \mathrm{d}h h (\ex{- \phi^{i\mathrm{s}}(h)/\kB T} - 1)
\label{eq:anderson}
\end{equation}
where $\kB$ is Boltzmann's constant, $T$ is the temperature, $\eta$ is the viscosity of the solution, and $\lambda_i$ is known as the Derjaguin length,\cite{derj47,ebbe12} which encodes the details of the non-contact interactions, and is typically of the order of a few angstroms. Overall attractive (resp.~repulsive) interactions lead to $\lambda_i^2>0$ ($\lambda_i^2<0$) and motion towards (away from) the substrate. Two cases of particular interest are the following: (i) If non-contact interactions with the substrate are dominated by the non-specific interactions which are present everywhere on the surface of the enzyme, we expect the phoretic responses of the free enzyme and the complex to be similar, with $\lambda_\mathrm{e} = \lambda_\mathrm{c} = \lambda$ and $\boldsymbol{v}_\mathrm{ph,e}=\boldsymbol{v}_\mathrm{ph,c}$, and the total phoretic velocity will simply be given by $\boldsymbol{V}_\mathrm{ph}  =  \frac{\kB T}{\eta} \lambda^2 \nabla c_\mathrm{s}$. (ii) If, on the other hand, the non-contact interactions are dominated by the attractive interactions near the binding pocket when the latter is free, we expect the free enzyme to show a much stronger phoretic response than the complex, with $|\lambda_\mathrm{e}| \gg |\lambda_\mathrm{c}|$ and $|\boldsymbol{v}_\mathrm{ph,e}| \gg |\boldsymbol{v}_\mathrm{ph,c}|$. In this case, the total phoretic velocity is given by $\boldsymbol{V}_\mathrm{ph} \simeq  \frac{\kB T}{\eta} \lambda_\mathrm{e}^2 \frac{K}{K + c_\mathrm{s}} \nabla c_\mathrm{s}$. It is interesting to note that the latter velocity has the same dependence on substrate concentration as the velocity that is derived from the thermodynamic theory in Ref~\citenum{schu13}.  

The second contribution to chemotaxis is due to binding-induced changes in the diffusion coefficient of the enzyme, with a velocity given by
\begin{eqnarray}
\boldsymbol{V}_\mathrm{bi}(\RR) =  -(D_\mathrm{c} - D_\mathrm{e}) \frac{K}{[K + c_\mathrm{s}(\RR]^2} \nabla c_\mathrm{s}(\RR).
\label{eq:Veff}
\end{eqnarray}
The velocity (\ref{eq:Veff}) is non-zero only if there is a gradient of substrate concentration \emph{and} the diffusion coefficient of the enzyme--substrate complex is different from the diffusion coefficient of the free enzyme. For substrates that \emph{enhance} diffusion, i.e.~with $D_\mathrm{c}>D_\mathrm{e}$, the velocity (\ref{eq:Veff}) points away from increasing substrate concentrations and thus leads to chemotaxis of the enzymes \emph{away from} higher concentrations of substrate. For substrates that \emph{inhibit} diffusion, i.e.~with $D_\mathrm{c}<D_\mathrm{e}$, the velocity (\ref{eq:Veff}) points towards increasing substrate concentrations and thus leads to chemotaxis of the enzymes \emph{towards} higher concentrations of substrate. This behavior is even more apparent when we notice that, in the absence of phoresis $\boldsymbol{V}_\mathrm{ph}(\RR) \equiv 0$, equation (\ref{eq:totalevol}) can be written as
\begin{eqnarray}
\partial_t c_\mathrm{e}^\mathrm{tot} (\RR;t) = \nabla^2 [D(\RR) c_\mathrm{e}^\mathrm{tot}]
\label{eq:totalevol2}
\end{eqnarray}
which implies that, in the absence of enzyme sources and sinks, the enzyme concentration will reach a zero-flux stationary profile $c_\mathrm{e}^\mathrm{tot} \propto 1/D(\RR)$, i.e.~will tend to concentrate in regions where its diffusion is slowest. The form (\ref{eq:totalevol2}) of the diffusion equation also implies that a phenomenological derivation of the correct diffusion equation would have only worked by using the It\^o treatment of the multiplicative noise, see Section~\ref{sec:pheno}. However, such a derivation would be neglecting the phoretic contribution to chemotaxis, which turns out to be very important.

Indeed, for a typical enzyme the non-contact interactions are expected to be attractive, leading to a phoretic velocity with $\lambda^2>0$. On the other hand, typical enzymes show enhanced diffusion, with $D_\mathrm{c}>D_\mathrm{e}$. As a consequence, phoresis typically points towards higher concentrations of substrate, whereas the contribution due to binding-induced enhanced diffusion points away from the substrate. Both contributions therefore compete against each other. Importantly, because the magnitude of the former decreases more slowly with increasing substrate concentration than the magnitude of the latter, imposing $|\boldsymbol{V}_\mathrm{ph}|=|\boldsymbol{V}_\mathrm{bi}|$ we find a \emph{critical substrate concentration} $c_\mathrm{s}^*$ above and below which phoresis and binding-induced enhanced diffusion dominate, respectively. For the particular case in which the Derjaguin lengths of the free enzyme and complex are similar, with $\lambda_\mathrm{e} = \lambda_\mathrm{c} = \lambda$, we find 
\begin{eqnarray}
c_\mathrm{s}^* = K \left( \sqrt{ \frac{|\alpha|}{6 \pi R_{\mathrm{e}} |\lambda^2| K} } - 1 \right)
\label{eq:critical}
\end{eqnarray}
 where we have defined the dimensionless change in diffusion $\alpha \equiv (D_\mathrm{c} - D_\mathrm{e})/D_\mathrm{e}$, and used the Stokes-Einstein relation $\kB T/\eta = 6 \pi D_\mathrm{e} R_\mathrm{e}$, where $R_\mathrm{e}$ is the hydrodynamic radius of the enzyme. In Ref~\citenum{agud18}, we showed that this competition can explain the conflicting experimental observations regarding whether urease chemotaxes towards\cite{seng13} or away from\cite{jee17} urea, the former experiments being dominated by phoresis, the latter by binding-induced enhanced diffusion. Furthermore, we verified that the rest of experimental observations of chemotaxis of enzymes towards their respective substrate are consistent with domination of the phoretic contribution, and showed that the observations of chemotaxis of molecular dyes towards a polymer to which they bind\cite{guha17} are consistent with domination of binding-induced \emph{inhibited} diffusion.

\begin{figure}
\centering
\includegraphics[width=0.7\linewidth]{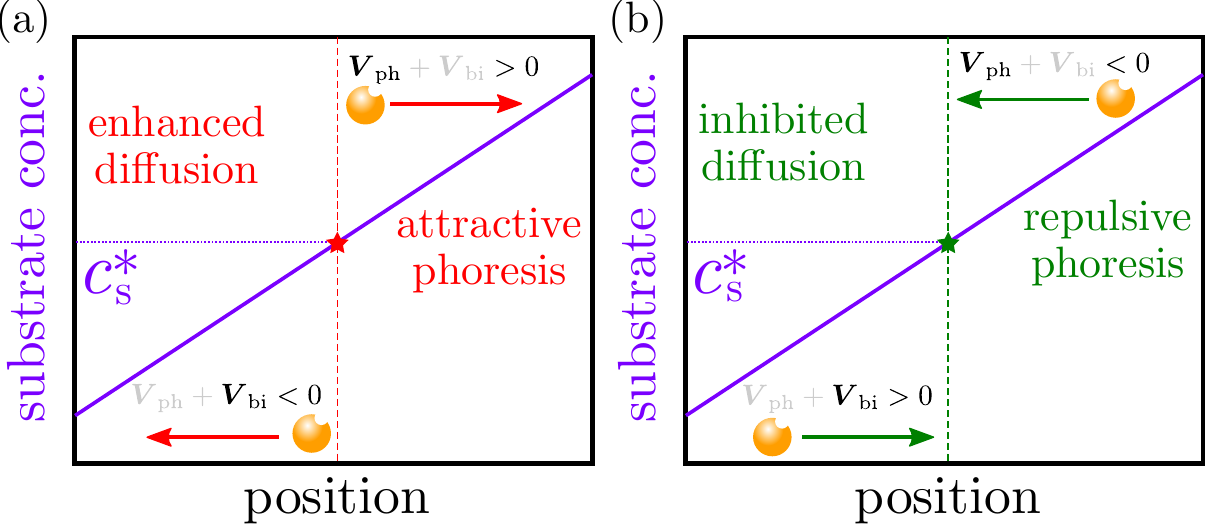}
\caption{Designing nanomachines that exploit the competition between phoresis and binding-induced changes in diffusion in an externally imposed substrate gradient. (a) Competition between attractive phoresis and enhanced diffusion leads to repulsion from regions with the critical substrate concentration $c_\mathrm{s}^*$. (b) Competition between repulsive phoresis and inhibited diffusion leads to accumulation in regions with the critical substrate concentration $c_\mathrm{s}^*$.}
   \label{fig:nanomachines}
\end{figure}

\subsection{Designing nanomachines with desirable response}

In principle, it should be possible to engineer enzyme-like nanomachines that show attractive or repulsive non-contact interactions with a substrate (e.g.~by tuning their surface chemistry); and that show enhanced or inhibited diffusion when specifically binding to the same substrate (e.g.~by tuning the mechanical properties of a modular nanomachine with internal degrees of freedom\cite{Illien2017b}).
When placed in a substrate gradient, a nanomachine displaying attractive phoresis ($\lambda^2>0$) combined with enhanced diffusion ($\alpha>0$), such as a typical enzyme, will move towards higher substrate concentrations in regions with $c_\mathrm{s}>c_\mathrm{s}^*$ (phoresis dominates), or toward lower substrate concentration in regions with $c_\mathrm{s}<c_\mathrm{s}^*$ (enhanced diffusion dominates). As a consequence, such a nanomachine will effectively be \emph{repelled} from regions with the critical substrate concentration $c_\mathrm{s}^*$, see Figure~\ref{fig:nanomachines}(a). Alternatively, a nanomachine displaying repulsive phoresis ($\lambda^2<0$) together with inhibited diffusion ($\alpha<0$) will move towards lower substrate concentrations in regions with $c_\mathrm{s}>c_\mathrm{s}^*$ (phoresis dominates), or toward higher substrate concentration in regions with $c_\mathrm{s}<c_\mathrm{s}^*$ (inhibited diffusion dominates), and as a consequence will be \emph{attracted} to regions with $c_\mathrm{s} = c_\mathrm{s}^*$, see Figure~\ref{fig:nanomachines}(b).

\section{Summary and outlook}

Although great progress has been achieved in recent years in order to understand the enhanced diffusion and chemotaxis of nanoscale enzymes and small molecules, much work remains to be done. In the case of enhanced diffusion, experimental evidence suggests that two different kinds of underlying mechanisms might be at play: on the one hand, there are active mechanisms that rely on the nonequilibrium catalytic step of the enzymatic reaction; on the other hand, there are equilibrium mechanisms caused by the binding and unbinding of substrate molecules to the enzyme. Whereas the equilibrium mechanisms are now relatively well identified,\cite{Illien2017a,Illien2017b} which precise active mechanisms are responsible for catalysis-induced enhanced diffusion is still unclear. Stochastic swimming or collective heating by the enzymes are possible candidates.\cite{Golestanian2015} More work, both experimental and theoretical, is needed to elucidate which type of mechanism will be important for a given enzyme.

Regarding enzyme chemotaxis, we have shown that many of the theoretical approaches put forward so far have shortcomings, due to relying on phenomenological approaches that link enhanced diffusion to chemotaxis without sufficient microscopic detail or an inaccurate implementation of hydrodynamics. We have introduced a microscopically-detailed theory of chemotaxis,\cite{agud18} which includes both non-contact interactions and specific binding between enzyme and substrate. The theory predicts two distinct contributions to chemotaxis, a competition between which can explain conflicting observations regarding the direction of enzyme chemotaxis at different substrate concentrations. We note that this theory does not depend on the catalytic activity or exothermicity of the enzyme. As in the case of enhanced diffusion, it is possible that active mechanisms may also play a role in chemotaxis under particular experimental conditions or for certain enzymes, but no clear candidates for such an active mechanism have been proposed so far.

The chemotactic capability of enzymes may be harnessed to construct drug delivery vehicles directed towards specific targets. Indeed, microparticles \cite{dey15} and vesicles \cite{jose17} decorated with enzymes have already been shown to exhibit directed motion in response to chemical gradients. The theory described above further implies that the existence of two competing chemotactic mechanisms may be exploited to construct nanomachines that are directed not just towards regions of highest or lowest concentration of a certain chemical, but towards or away from regions with a \emph{specific} chemical composition.

We note that we have focused here on theories valid for a single enzyme, owing to the fact that experiments are typically performed at dilute enzyme concentrations. At higher concentrations, however, collective effects arising from the interaction between many enzymes, both through their substrate and product concentration fields, as well as through hydrodynamic interactions, are expected to become important. Such a situation, particularly in cases where the product of one enzyme is the substrate of another (thus forming an enzymatic cascade) should be directly relevant to metabolon formation in metabolic pathways. \cite{Wu2015,zhao17} In Ref~\citenum{saha14}, the collective effects that arise when many catalytic colloidal swimmers interact with each other were studied in detail. Complex patterns including cluster and aster formation, as well as collective oscillations were found. It is an exciting prospect to think that such collective behaviour might arise for nanoscale enzymes as well.

\section*{Biographical Information}

Jaime Agudo-Canalejo obtained his undergraduate degree in Physics from the Complutense University of Madrid (2012), and his Ph.D.~from the Technical University of Berlin (2016). His doctoral work was conducted at the Max Planck Institute of Colloids and Interfaces, where he also held a postdoctoral position until 2017. Since 2017, he is a postdoctoral fellow jointly at the University of Oxford
and at Penn State University.
He is interested in the theoretical modelling of soft and active matter, with particular focus on biological systems.

Pierre Illien completed his undergraduate studies at Ecole Normale Sup\'erieure (Paris), and obtained his PhD in theoretical physics from Universit\'e-Pierre-et-Marie-Curie (Paris) in 2015. From 2015 to 2017, he was a postdoctoral researcher
at the University of Oxford and at Penn State University.
He recently joined ESPCI Paris as a postdoctoral fellow. His research interests include theoretical approaches to nonequilibrium statistical mechanics and their applications to soft and biological matter.

Tunrayo Adeleke-Larodo has been a D. Phil student
at the University of Oxford
since 2016. Previously she was an undergraduate in Mathematical Physics at the University of Edinburgh. Her current research is in statistical physics, specifically of non-equilibrium and active matter systems.

Ramin Golestanian is director at the Max Planck Institute for Dynamics and Self-Organization in G\"ottingen and Professor of Theoretical Condensed Matter Physics at the University of Oxford. He obtained his BSc from Sharif University of Technology in Tehran, and his MSc and PhD from the Institute for Advanced Studies in Basic Sciences (IASBS) in Zanjan. After an independent postdoctoral fellowship at the Kavli Institute for Theoretical Physics at the University of California at Santa Barbara, he held academic positions at IASBS, the University of Sheffield, and Oxford University. He has a broad interest in various aspects of nonequilibrium statistical physics, soft matter, and biological physics, and is distinguished for his work on active matter, and in particular, for his role in developing microscopic swimmers and active colloids.

\begin{acknowledgement}

J.A-C., P.I., and R.G.~were supported by the US National Science Foundation under MRSEC Grant number DMR-1420620. T.A-L. acknowledges the support of EPSRC.

\end{acknowledgement}

%
%
%

\bibliography{biblio}

\providecommand{\latin}[1]{#1}
\makeatletter
\providecommand{\doi}
  {\begingroup\let\do\@makeother\dospecials
  \catcode`\{=1 \catcode`\}=2 \doi@aux}
\providecommand{\doi@aux}[1]{\endgroup\texttt{#1}}
\makeatother
\providecommand*\mcitethebibliography{\thebibliography}
\csname @ifundefined\endcsname{endmcitethebibliography}
  {\let\endmcitethebibliography\endthebibliography}{}
\begin{mcitethebibliography}{35}
\providecommand*\natexlab[1]{#1}
\providecommand*\mciteSetBstSublistMode[1]{}
\providecommand*\mciteSetBstMaxWidthForm[2]{}
\providecommand*\mciteBstWouldAddEndPuncttrue
  {\def\EndOfBibitem{\unskip.}}
\providecommand*\mciteBstWouldAddEndPunctfalse
  {\let\EndOfBibitem\relax}
\providecommand*\mciteSetBstMidEndSepPunct[3]{}
\providecommand*\mciteSetBstSublistLabelBeginEnd[3]{}
\providecommand*\EndOfBibitem{}
\mciteSetBstSublistMode{f}
\mciteSetBstMaxWidthForm{subitem}{(\alph{mcitesubitemcount})}
\mciteSetBstSublistLabelBeginEnd
  {\mcitemaxwidthsubitemform\space}
  {\relax}
  {\relax}

\bibitem[Dey and Sen(2017)Dey, and Sen]{dey17}
Dey,~K.~K.; Sen,~A. {Chemically Propelled Molecules and Machines}. \emph{J. Am.
  Chem. Soc.} \textbf{2017}, \emph{139}, 7666--7676\relax
\mciteBstWouldAddEndPuncttrue
\mciteSetBstMidEndSepPunct{\mcitedefaultmidpunct}
{\mcitedefaultendpunct}{\mcitedefaultseppunct}\relax
\EndOfBibitem
\bibitem[M{\o}ller(2010)]{moll10}
M{\o}ller,~B.~L. {Dynamic metabolons}. \emph{Science} \textbf{2010},
  \emph{330}, 1328--1329\relax
\mciteBstWouldAddEndPuncttrue
\mciteSetBstMidEndSepPunct{\mcitedefaultmidpunct}
{\mcitedefaultendpunct}{\mcitedefaultseppunct}\relax
\EndOfBibitem
\bibitem[Zhao \latin{et~al.}(2018)Zhao, Palacci, Yadav, Spiering, Gilson,
  Butler, Hess, Benkovic, and Sen]{zhao17}
Zhao,~X.; Palacci,~H.; Yadav,~V.; Spiering,~M.~M.; Gilson,~M.~K.;
  Butler,~P.~J.; Hess,~H.; Benkovic,~S.~J.; Sen,~A. {Substrate-driven
  chemotactic assembly in an enzyme cascade}. \emph{Nat. Chem.} \textbf{2018},
  \emph{10}, 311--317\relax
\mciteBstWouldAddEndPuncttrue
\mciteSetBstMidEndSepPunct{\mcitedefaultmidpunct}
{\mcitedefaultendpunct}{\mcitedefaultseppunct}\relax
\EndOfBibitem
\bibitem[Wu \latin{et~al.}(2015)Wu, Pelster, and Minteer]{Wu2015}
Wu,~F.; Pelster,~L.~N.; Minteer,~S.~D. {Krebs cycle metabolon formation:
  metabolite concentration gradient enhanced compartmentation of sequential
  enzymes.} \emph{Chem. Commun.} \textbf{2015}, \emph{51}, 1244--1247\relax
\mciteBstWouldAddEndPuncttrue
\mciteSetBstMidEndSepPunct{\mcitedefaultmidpunct}
{\mcitedefaultendpunct}{\mcitedefaultseppunct}\relax
\EndOfBibitem
\bibitem[Schwander \latin{et~al.}(2016)Schwander, {Schada von Borzyskowski},
  Burgener, Cortina, and Erb]{schw16}
Schwander,~T.; {Schada von Borzyskowski},~L.; Burgener,~S.; Cortina,~N.~S.;
  Erb,~T.~J. {A synthetic pathway for the fixation of carbon dioxide in vitro.}
  \emph{Science} \textbf{2016}, \emph{354}, 900--904\relax
\mciteBstWouldAddEndPuncttrue
\mciteSetBstMidEndSepPunct{\mcitedefaultmidpunct}
{\mcitedefaultendpunct}{\mcitedefaultseppunct}\relax
\EndOfBibitem
\bibitem[Muddana \latin{et~al.}(2010)Muddana, Sengupta, Mallouk, Sen, and
  Butler]{Muddana2010a}
Muddana,~H.~S.; Sengupta,~S.; Mallouk,~T.~E.; Sen,~A.; Butler,~P.~J. {Substrate
  catalysis enhances single-enzyme diffusion}. \emph{J. Am. Chem. Soc.}
  \textbf{2010}, \emph{132}, 2110--2111\relax
\mciteBstWouldAddEndPuncttrue
\mciteSetBstMidEndSepPunct{\mcitedefaultmidpunct}
{\mcitedefaultendpunct}{\mcitedefaultseppunct}\relax
\EndOfBibitem
\bibitem[Sengupta \latin{et~al.}(2013)Sengupta, Dey, Muddana, Tabouillot,
  Ibele, Butler, and Sen]{seng13}
Sengupta,~S.; Dey,~K.~K.; Muddana,~H.~S.; Tabouillot,~T.; Ibele,~M.~E.;
  Butler,~P.~J.; Sen,~A. {Enzyme molecules as nanomotors}. \emph{J. Am. Chem.
  Soc.} \textbf{2013}, \emph{135}, 1406--1414\relax
\mciteBstWouldAddEndPuncttrue
\mciteSetBstMidEndSepPunct{\mcitedefaultmidpunct}
{\mcitedefaultendpunct}{\mcitedefaultseppunct}\relax
\EndOfBibitem
\bibitem[Sengupta \latin{et~al.}(2014)Sengupta, Spiering, Dey, Duan, Patra,
  Butler, Astumian, Benkovic, and Sen]{seng14}
Sengupta,~S.; Spiering,~M.~M.; Dey,~K.~K.; Duan,~W.; Patra,~D.; Butler,~P.~J.;
  Astumian,~R.~D.; Benkovic,~S.~J.; Sen,~A. {DNA polymerase as a molecular
  motor and pump}. \emph{ACS Nano} \textbf{2014}, \emph{8}, 2410--2418\relax
\mciteBstWouldAddEndPuncttrue
\mciteSetBstMidEndSepPunct{\mcitedefaultmidpunct}
{\mcitedefaultendpunct}{\mcitedefaultseppunct}\relax
\EndOfBibitem
\bibitem[Riedel \latin{et~al.}(2014)Riedel, Gabizon, Wilson, Hamadani,
  Tsekouras, Marqusee, Presse, and Bustamante]{Riedel2015}
Riedel,~C.; Gabizon,~R.; Wilson,~C. A.~M.; Hamadani,~K.; Tsekouras,~K.;
  Marqusee,~S.; Presse,~S.; Bustamante,~C. {The heat released during catalytic
  turnover enhances the diffusion of an enzyme}. \emph{Nature} \textbf{2014},
  \emph{517}, 227--230\relax
\mciteBstWouldAddEndPuncttrue
\mciteSetBstMidEndSepPunct{\mcitedefaultmidpunct}
{\mcitedefaultendpunct}{\mcitedefaultseppunct}\relax
\EndOfBibitem
\bibitem[Illien \latin{et~al.}(2017)Illien, Zhao, Dey, Butler, Sen, and
  Golestanian]{Illien2017a}
Illien,~P.; Zhao,~X.; Dey,~K.~K.; Butler,~P.~J.; Sen,~A.; Golestanian,~R.
  {Exothermicity is not a necessary condition for enhanced diffusion of
  enzymes}. \emph{Nano Lett.} \textbf{2017}, \emph{17}, 4415--4420\relax
\mciteBstWouldAddEndPuncttrue
\mciteSetBstMidEndSepPunct{\mcitedefaultmidpunct}
{\mcitedefaultendpunct}{\mcitedefaultseppunct}\relax
\EndOfBibitem
\bibitem[Yu \latin{et~al.}(2009)Yu, Jo, Kounovsky, Pablo, and Schwartz]{yu09}
Yu,~H.; Jo,~K.; Kounovsky,~K.~L.; Pablo,~J. J.~D.; Schwartz,~D.~C. {Molecular
  propulsion: Chemical sensing and chemotaxis of DNA driven by RNA polymerase}.
  \emph{J. Am. Chem. Soc.} \textbf{2009}, \emph{131}, 5722--5723\relax
\mciteBstWouldAddEndPuncttrue
\mciteSetBstMidEndSepPunct{\mcitedefaultmidpunct}
{\mcitedefaultendpunct}{\mcitedefaultseppunct}\relax
\EndOfBibitem
\bibitem[Dey \latin{et~al.}(2014)Dey, Das, Poyton, Sengupta, Butler, Cremer,
  and Sen]{dey14}
Dey,~K.~K.; Das,~S.; Poyton,~M.~F.; Sengupta,~S.; Butler,~P.~J.; Cremer,~P.~S.;
  Sen,~A. {Chemotactic separation of enzymes}. \emph{ACS Nano} \textbf{2014},
  \emph{8}, 11941--11949\relax
\mciteBstWouldAddEndPuncttrue
\mciteSetBstMidEndSepPunct{\mcitedefaultmidpunct}
{\mcitedefaultendpunct}{\mcitedefaultseppunct}\relax
\EndOfBibitem
\bibitem[Guha \latin{et~al.}(2017)Guha, Mohajerani, Collins, Ghosh, Sen, and
  Velegol]{guha17}
Guha,~R.; Mohajerani,~F.; Collins,~M.; Ghosh,~S.; Sen,~A.; Velegol,~D.
  {Chemotaxis of Molecular Dyes in Polymer Gradients in Solution}. \emph{J. Am.
  Chem. Soc.} \textbf{2017}, \emph{139}, 15588--15591\relax
\mciteBstWouldAddEndPuncttrue
\mciteSetBstMidEndSepPunct{\mcitedefaultmidpunct}
{\mcitedefaultendpunct}{\mcitedefaultseppunct}\relax
\EndOfBibitem
\bibitem[Jee \latin{et~al.}(2018)Jee, Dutta, Cho, Tlusty, and Granick]{jee17}
Jee,~A.-Y.; Dutta,~S.; Cho,~Y.-K.; Tlusty,~T.; Granick,~S. {Enzyme leaps fuel
  antichemotaxis}. \emph{Proc. Natl. Acad. Sci. U. S. A.} \textbf{2018},
  \emph{115}, 14--18\relax
\mciteBstWouldAddEndPuncttrue
\mciteSetBstMidEndSepPunct{\mcitedefaultmidpunct}
{\mcitedefaultendpunct}{\mcitedefaultseppunct}\relax
\EndOfBibitem
\bibitem[Illien \latin{et~al.}(2017)Illien, Adeleke-Larodo, and
  Golestanian]{Illien2017b}
Illien,~P.; Adeleke-Larodo,~T.; Golestanian,~R. {Diffusion of an enzyme: The
  role of fluctuation-induced hydrodynamic coupling}. \emph{Europhys. Lett.}
  \textbf{2017}, \emph{119}, 40002\relax
\mciteBstWouldAddEndPuncttrue
\mciteSetBstMidEndSepPunct{\mcitedefaultmidpunct}
{\mcitedefaultendpunct}{\mcitedefaultseppunct}\relax
\EndOfBibitem
\bibitem[Agudo-Canalejo \latin{et~al.}(2018)Agudo-Canalejo, Illien, and
  Golestanian]{agud18}
Agudo-Canalejo,~J.; Illien,~P.; Golestanian,~R. {Phoresis and Enhanced
  Diffusion Compete in Enzyme Chemotaxis}. \emph{Nano Lett.} \textbf{2018},
  \emph{18}, 2711--2717\relax
\mciteBstWouldAddEndPuncttrue
\mciteSetBstMidEndSepPunct{\mcitedefaultmidpunct}
{\mcitedefaultendpunct}{\mcitedefaultseppunct}\relax
\EndOfBibitem
\bibitem[Golestanian(2015)]{Golestanian2015}
Golestanian,~R. {Enhanced diffusion of enzymes that catalyze exothermic
  reactions}. \emph{Phys. Rev. Lett.} \textbf{2015}, \emph{115}, 108102\relax
\mciteBstWouldAddEndPuncttrue
\mciteSetBstMidEndSepPunct{\mcitedefaultmidpunct}
{\mcitedefaultendpunct}{\mcitedefaultseppunct}\relax
\EndOfBibitem
\bibitem[Dey \latin{et~al.}(2015)Dey, Zhao, Tansi, Mendez-Ortiz,
  Cordova-Figueroa, Golestanian, and Sen]{dey15}
Dey,~K.~K.; Zhao,~X.; Tansi,~B.~M.; Mendez-Ortiz,~W.~J.;
  Cordova-Figueroa,~U.~M.; Golestanian,~R.; Sen,~A. {Micromotors Powered by
  Enzyme Catalysis}. \emph{Nano Lett.} \textbf{2015}, \emph{15},
  8311--8315\relax
\mciteBstWouldAddEndPuncttrue
\mciteSetBstMidEndSepPunct{\mcitedefaultmidpunct}
{\mcitedefaultendpunct}{\mcitedefaultseppunct}\relax
\EndOfBibitem
\bibitem[Bai and Wolynes(2015)Bai, and Wolynes]{Bai2015}
Bai,~X.; Wolynes,~P.~G. {On the hydrodynamics of swimming enzymes}. \emph{J.
  Chem. Phys.} \textbf{2015}, \emph{143}, 165101\relax
\mciteBstWouldAddEndPuncttrue
\mciteSetBstMidEndSepPunct{\mcitedefaultmidpunct}
{\mcitedefaultendpunct}{\mcitedefaultseppunct}\relax
\EndOfBibitem
\bibitem[Mikhailov and Kapral(2015)Mikhailov, and Kapral]{Mikhailov2015}
Mikhailov,~A.~S.; Kapral,~R. {Hydrodynamic collective effects of active protein
  machines in solution and lipid bilayers}. \emph{Proc. Natl. Acad. Sci. U. S.
  A.} \textbf{2015}, \emph{112}, E3639--E3644\relax
\mciteBstWouldAddEndPuncttrue
\mciteSetBstMidEndSepPunct{\mcitedefaultmidpunct}
{\mcitedefaultendpunct}{\mcitedefaultseppunct}\relax
\EndOfBibitem
\bibitem[Rago \latin{et~al.}(2015)Rago, Saltzberg, Allen, and Tolan]{Rago2015}
Rago,~F.; Saltzberg,~D.; Allen,~K.~N.; Tolan,~D.~R. {Enzyme Substrate
  Specificity Conferred by Distinct Conformational Pathways}. \emph{J. Am.
  Chem. Soc.} \textbf{2015}, \emph{137}, 13876--13886\relax
\mciteBstWouldAddEndPuncttrue
\mciteSetBstMidEndSepPunct{\mcitedefaultmidpunct}
{\mcitedefaultendpunct}{\mcitedefaultseppunct}\relax
\EndOfBibitem
\bibitem[Roberts \latin{et~al.}(2012)Roberts, Miller, Roitberg, and
  Merz]{Roberts2012}
Roberts,~B.~P.; Miller,~B.~R.; Roitberg,~A.~E.; Merz,~K.~M. {Wide-Open Flaps
  Are Key to Urease Activity}. \emph{J. Am. Chem. Soc.} \textbf{2012},
  \emph{134}, 9934--9937\relax
\mciteBstWouldAddEndPuncttrue
\mciteSetBstMidEndSepPunct{\mcitedefaultmidpunct}
{\mcitedefaultendpunct}{\mcitedefaultseppunct}\relax
\EndOfBibitem
\bibitem[Joseph \latin{et~al.}(2017)Joseph, Contini, Cecchin, Nyberg,
  Ruiz-Perez, Gaitzsch, Fullstone, Tian, Azizi, Preston, Volpe, and
  Battaglia]{jose17}
Joseph,~A.; Contini,~C.; Cecchin,~D.; Nyberg,~S.; Ruiz-Perez,~L.; Gaitzsch,~J.;
  Fullstone,~G.; Tian,~X.; Azizi,~J.; Preston,~J.; Volpe,~G.; Battaglia,~G.
  {Chemotactic synthetic vesicles: Design and applications in blood-brain
  barrier crossing}. \emph{Sci. Adv.} \textbf{2017}, \emph{3}, e1700362\relax
\mciteBstWouldAddEndPuncttrue
\mciteSetBstMidEndSepPunct{\mcitedefaultmidpunct}
{\mcitedefaultendpunct}{\mcitedefaultseppunct}\relax
\EndOfBibitem
\bibitem[Schurr \latin{et~al.}(2013)Schurr, Fujimoto, Huynh, and Chiu]{schu13}
Schurr,~J.~M.; Fujimoto,~B.~S.; Huynh,~L.; Chiu,~D.~T. {A theory of
  macromolecular chemotaxis}. \emph{J. Phys. Chem. B} \textbf{2013},
  \emph{117}, 7626--7652\relax
\mciteBstWouldAddEndPuncttrue
\mciteSetBstMidEndSepPunct{\mcitedefaultmidpunct}
{\mcitedefaultendpunct}{\mcitedefaultseppunct}\relax
\EndOfBibitem
\bibitem[Anderson(1989)]{ande89}
Anderson,~J.~L. {Colloid transport by interfacial forces}. \emph{Annu. Rev.
  Fluid Mech.} \textbf{1989}, \emph{21}, 61--99\relax
\mciteBstWouldAddEndPuncttrue
\mciteSetBstMidEndSepPunct{\mcitedefaultmidpunct}
{\mcitedefaultendpunct}{\mcitedefaultseppunct}\relax
\EndOfBibitem
\bibitem[Hong \latin{et~al.}(2007)Hong, Blackman, Kopp, Sen, and
  Velegol]{hong07}
Hong,~Y.; Blackman,~N. M.~K.; Kopp,~N.~D.; Sen,~A.; Velegol,~D. {Chemotaxis of
  nonbiological colloidal rods}. \emph{Phys. Rev. Lett.} \textbf{2007},
  \emph{99}, 178103\relax
\mciteBstWouldAddEndPuncttrue
\mciteSetBstMidEndSepPunct{\mcitedefaultmidpunct}
{\mcitedefaultendpunct}{\mcitedefaultseppunct}\relax
\EndOfBibitem
\bibitem[Butler \latin{et~al.}(2015)Butler, Dey, and Sen]{butl15}
Butler,~P.~J.; Dey,~K.~K.; Sen,~A. {Impulsive Enzymes: A New Force in
  Mechanobiology}. \emph{Cell. Mol. Bioeng.} \textbf{2015}, \emph{8},
  106--118\relax
\mciteBstWouldAddEndPuncttrue
\mciteSetBstMidEndSepPunct{\mcitedefaultmidpunct}
{\mcitedefaultendpunct}{\mcitedefaultseppunct}\relax
\EndOfBibitem
\bibitem[Weistuch and Press{\'{e}}(2017)Weistuch, and Press{\'{e}}]{weis17}
Weistuch,~C.; Press{\'{e}},~S. {Spatiotemporal Organization of Catalysts Driven
  by Enhanced Diffusion}. \emph{J. Phys. Chem. B} \textbf{2017}, \emph{122},
  5286--5290\relax
\mciteBstWouldAddEndPuncttrue
\mciteSetBstMidEndSepPunct{\mcitedefaultmidpunct}
{\mcitedefaultendpunct}{\mcitedefaultseppunct}\relax
\EndOfBibitem
\bibitem[Lau and Lubensky(2007)Lau, and Lubensky]{lau07}
Lau,~A. W.~C.; Lubensky,~T.~C. {State-dependent diffusion: Thermodynamic
  consistency and its path integral formulation}. \emph{Phys. Rev. E}
  \textbf{2007}, \emph{76}, 011123\relax
\mciteBstWouldAddEndPuncttrue
\mciteSetBstMidEndSepPunct{\mcitedefaultmidpunct}
{\mcitedefaultendpunct}{\mcitedefaultseppunct}\relax
\EndOfBibitem
\bibitem[van Kampen(1981)]{vanKampen1981}
van Kampen,~N.~G. It{\^o} versus Stratonovich. \emph{J. Stat. Phys.}
  \textbf{1981}, \emph{24}, 175--187\relax
\mciteBstWouldAddEndPuncttrue
\mciteSetBstMidEndSepPunct{\mcitedefaultmidpunct}
{\mcitedefaultendpunct}{\mcitedefaultseppunct}\relax
\EndOfBibitem
\bibitem[Schnitzer(1993)]{Schnitzer1993}
Schnitzer,~M.~J. {Theory of continuum random walks and application to
  chemotaxis}. \emph{Phys. Rev. E} \textbf{1993}, \emph{48}, 2553--2568\relax
\mciteBstWouldAddEndPuncttrue
\mciteSetBstMidEndSepPunct{\mcitedefaultmidpunct}
{\mcitedefaultendpunct}{\mcitedefaultseppunct}\relax
\EndOfBibitem
\bibitem[Derjaguin \latin{et~al.}(1947)Derjaguin, Sidorenkov, Zubashchenkov,
  and Kiseleva]{derj47}
Derjaguin,~B.~V.; Sidorenkov,~G.~P.; Zubashchenkov,~E.~A.; Kiseleva,~E.~V.
  {Kinetic phenomena in boundary films of liquids}. \emph{Kolloidn. Zh}
  \textbf{1947}, \emph{9}, 335--347\relax
\mciteBstWouldAddEndPuncttrue
\mciteSetBstMidEndSepPunct{\mcitedefaultmidpunct}
{\mcitedefaultendpunct}{\mcitedefaultseppunct}\relax
\EndOfBibitem
\bibitem[Ebbens \latin{et~al.}(2012)Ebbens, Tu, Howse, and Golestanian]{ebbe12}
Ebbens,~S.; Tu,~M.~H.; Howse,~J.~R.; Golestanian,~R. {Size dependence of the
  propulsion velocity for catalytic Janus-sphere swimmers}. \emph{Phys. Rev. E}
  \textbf{2012}, \emph{85}, 020401(R)\relax
\mciteBstWouldAddEndPuncttrue
\mciteSetBstMidEndSepPunct{\mcitedefaultmidpunct}
{\mcitedefaultendpunct}{\mcitedefaultseppunct}\relax
\EndOfBibitem
\bibitem[Saha \latin{et~al.}(2014)Saha, Golestanian, and Ramaswamy]{saha14}
Saha,~S.; Golestanian,~R.; Ramaswamy,~S. {Clusters, asters, and collective
  oscillations in chemotactic colloids}. \emph{Phys. Rev. E} \textbf{2014},
  \emph{89}, 062316\relax
\mciteBstWouldAddEndPuncttrue
\mciteSetBstMidEndSepPunct{\mcitedefaultmidpunct}
{\mcitedefaultendpunct}{\mcitedefaultseppunct}\relax
\EndOfBibitem
\end{mcitethebibliography}

\end{document}